\documentclass[english,pra, twocolumn, showpacs, showkeyw]{revtex4-1}
\usepackage[T1]{fontenc}
\usepackage{xcolor}
\usepackage{amsmath}
\usepackage{graphicx}

\makeatletter
\usepackage[colorlinks = true,
            linkcolor =red,
            urlcolor  =magenta,
            citecolor = blue,
            anchorcolor = blue]{hyperref}

\@ifundefined{showcaptionsetup}{}{%
 \PassOptionsToPackage{caption=false}{subfig}}
\usepackage{subfig}
\makeatother

\usepackage{babel}
\begin{document}
\title{Single-Photon-Subtracted-Squeezed-Vacuum-State Based Postselected
Weak Measurement and its Applications}
\author{Janarbek Yuanbek$^{1}$Akbar Islam\textsuperscript{1} Ahmad Abliz$^{1}$}
\email{aahmad@126.com}

\author{Yusuf Turek$^{2}$}
\email{yusuftu1984@hotmail.com}

\address{$^{1}$School of Physics and Electronic Engineering, Xinjiang Normal
University, Urumqi, Xinjiang 830054, China}
\address{$^{2}$School of Physics, Liaoning University, Shenyang, Liaoning
110036, China}
\date{\today}
\begin{abstract}
In this paper, we study the\textcolor{brown}{{} }effects of postselected
von Neumann measurement on the nonclassicality of the Single-Photon-Subtracted-Squeezed-Vacuum-State
(SPSSVS). We calculate the squeezing effect, Mandel factor, Wigner
function, signal-to-noise ratio (SNR)s and state distance function.
We found that postselected von Neumann measurement has positive effects
on the optimization of SPSSVS. In particular, by properly choosing
the anomalous weak value, the nonclassical inherent features of SPSSVS
such as squeezing, photon statistics and phase space distribution
can be optimized significantly. The advantages of postselected weak
measurement on improving the SNR compared to non-postselected measurement
scheme is also confirmed. The superiority of SPSSVS based postselected
weak measurement in quantum state optimization may have potential
applications of in the associated quantum information processing.
\end{abstract}
\maketitle

\section{INTRODUCTION\label{sec:1 }}

As is well known, the information of a microscopic system is contained
in its quantum state. Therefore, it is of utmost importance to find
the quantum state that describes a microscopic system. While any quantum
system can be described by its corresponding quantum state, not all
states exhibit good nonclassical properties.\textcolor{red}{{} }The
quantum weak measurements proposed by physicists such as Aharonov,
Albert, and Vaidman in the late 1980s not only generalize the original
measurement theory but also provide a good solution for the development
of quantum control \citep{PhysRevLett.60.1351}. In a single weak
measurement, the information obtained from the measured system by
the measuring instrument is limited. Nevertheless, the partially collapsed
quantum state can still retain some information through repetitive
measurements, thereby significantly increasing the measurement accuracy.
Using the weak signal amplifying principle, the weak measurement method
also amplifies measurement results with large values, even in the
weak coupling regime. In quantum optics, photon number statistics
are used to describe the nonclassical properties of radiated light
fields. A significant amount of theoretical and experimental work
has been conducted in this area to confirm the effectiveness of squeezing
effects and phase-space distribution functions. The preparation and
optimisation of non-classical properties of the quantum state are
vital in quantum information processing, including the generation
and detection of single photons \citep{RN1927,Zhang2014}, quantum
computing \citep{RN1928,Chuang2010}, quantum teleportation \citep{PhysRevA.54.2614,PhysRevA.54.2629,PhysRevA.60.937,PhysRevA.67.033802,PhysRevA.67.042314,PhysRevA.68.052308,PhysRevD.23.1693},
generation and manipulation of atom-light entanglement and precision
measurements \citep{Hacker2018DeterministicCO}. However, the realisation
of these processes depends on the generating, analysing and optimisation
of the relevant quantum states, such as the coherent state \citep{PhysRev.131.2766,PhysRev.140.B676,PhysRevD.4.2309},
squeezing state \citep{Carranza:12,Andersen201530YO,RN1932}, photon
number states \citep{PhysRevLett.56.58,PhysRevA.36.4547,PhysRevA.39.3414,RN1930,Liu_2004,Waks2006,RN1931},
even and odd coherent states \citep{PhysRevLett.75.4011,PhysRevA.13.2226}.
However, with the emerge of state-of-the-art technologies, the above
mentioned states can no longer meet the needs of practical applications
in quantum information science. The construction of new quantum states
has become a new pursuit for researchers in related fields. In recent
years, various schemes have been proposed for creating new quantum
states and their related properties have been investigated to some
extent\citep{2011JMOp...58..890A}.

In this manuscript, we study the application of post-selective measurement
theory to precision measurements using the SPSSVS \citep{PhysRevA.75.032104,PhysRevA.43.492}
as a pointer. In order to investigate the advantages of the SPSSVS-based
post-selective weak measurements in precision measurements, we present
both the analytical expressions and numerical results including the
second-order correlation function, Mandel factor, squeezing effect,
Wigner function, SNR and state distance. The results show that the
SPSSVS exhibits weaker statistical properties of the sub-Poisson photon
distribution. Theoretical analyses show that the post-selective measurement
theory with the SPSSVS as a pointer state does have weaker nonclassical
properties than the original state{} \citep{2021Single}.

This paper is organized as follows. In Sec.\ref{sec:2}, we introduce
the main concepts of our scheme and the final pointer state after
the postselected measurement, which will be used throughout the study,
then we give details of the squeezing effects of the final pointer
state by using the definition of squeezing. In the subsequent sections,
we study Mandel factor and Wigner function of the SPSSVS. In Sec.\ref{sec:5}
we calculate the SNR of SPSSVS after postselected measurement. In
Sec.\ref{sec:6} state distance is analyzed in order to examine the
effects of postselected measurement on the initial state. In Sec.\ref{sec:7}
we present our conlusion and remarks.

\section{The effects of postselected measurement on the properties of SPSSVS
\label{sec:2 }}

In measurement theory, the total Hamiltonian has a general form of
consisting three parts as $\hat{H}=\hat{H}_{s}+\hat{H}_{p}+\hat{H}_{int}$.
Here, $\hat{H}_{s}$ and $\hat{H}_{p}$ represent the Hamiltonians
of the measured system and pointer (mesuring device), respectively,
and $\hat{H}_{int}$ is the interaction Hamiltonian. According to
the ideal measurement theory \citep{vonNeumann+2018}, the explicit
expressions of the Hamiltonian for the pointer and measured system
do not affect the measurement results. This interaction Hamiltonian
contains the main information about the pointer and the measured system.
In this work we take the interaction Hamiltonian as

\begin{equation}
\hat{H}_{int}=g\delta(t-t_{0})\hat{\sigma}_{x}\otimes\hat{P}.\label{eq:1}
\end{equation}
The function $g\delta(t-t_{0})$ represents the interaction coupling
between the pointer and the measured system, and $\int_{0}^{t}g\delta(t-t_{0})d\tau=g$.
$\hat{\sigma_{x}}=\vert\uparrow_{x}\rangle\langle\uparrow_{x}\vert-\vert\downarrow_{x}\rangle\langle\downarrow_{x}\vert$
is the Pauli $x$ operator of the system to be measured and $\vert\uparrow_{x}\rangle$
and $\vert\downarrow_{x}\rangle$ are the eigenstates of $\hat{\sigma}_{x}$
with corresponding eigenvalues $1$ and $-1$, respectively. In the
above Hamiltonian the $\hat{P}$ denotes the momentum operator of
the pointer which is the conjugate of $\hat{X}$ and $\left[\hat{X},\hat{P}\right]=i\hat{I}$.
Our goal in this work is to investigate the effects of post-selected
measurement on the inherent properties of the SPSSVS. Thus, in the
present work, we take the polarization and spatial degrees of freedom
of the SPSSVS as the measured system and pointer, respectively. We
assume that the initial state of the total system is prepared as $\vert\psi_{i}\rangle\otimes\vert\phi\rangle$.
Here $\vert\psi_{i}\rangle=\cos\frac{\alpha}{2}\vert\uparrow_{z}\rangle+e^{i\delta}\sin\frac{\alpha}{2}\vert\downarrow_{z}\rangle$
with $\delta\in[0,2\pi]$ and $\alpha\in[0,\pi)$, $\vert\phi\rangle$
is the SPSSVS defined as 
\begin{equation}
\begin{aligned}\vert\phi\rangle & =\kappa\hat{a}\vert\xi\rangle=\kappa\hat{a}\hat{S}(\xi)\vert0\rangle.\end{aligned}
\label{eq:2}
\end{equation}
Here, $\kappa=\sinh^{-1}r$ and $\hat{S}(\xi)=\exp[\frac{1}{2}\xi\hat{a}^{\dagger2}-\frac{1}{2}\xi^{\ast}\hat{a}^{2}],\xi=re^{i\theta}$is
the squeezing operator. The initial state $\vert\psi_{i}\rangle$
can be prepared in the optical lab by using quarter and half wave
plates. Under the unitary evolution operator $\hat{U}(t)=\exp\left(-i\int_{0}^{t}\hat{H}_{int}d\tau\right)$,
the state $\vert\psi_{i}\rangle\otimes\vert\phi\rangle$evolves to

\begin{align}
\vert\Psi\rangle & =e^{-ig\hat{\sigma}_{x}\otimes\hat{P}_{x}}\vert\psi_{i}\rangle\otimes\vert\phi\rangle\nonumber \\
 & =\frac{1}{2}\left[\!(\hat{I}+\hat{\sigma}_{x})\otimes\hat{D}\left(\frac{s}{2}\right)+\!\!(\hat{I}-\!\!\hat{\sigma}_{x})\otimes\hat{D}\left(\!-\frac{s}{2}\right)\right]\!\vert\psi_{i}\rangle\!\otimes\!\vert\phi\rangle.\!\!\!\label{eq:3}
\end{align}
In the derivation of the above expression we have written the momentum
operator $\hat{P}$ in terms of the annihilation and creation operator,
$\hat{a}$ and $\hat{a}^{\dagger}$, as $\hat{P_{x}}=\frac{i}{2\sigma}\left(\hat{a}^{\dagger}-\hat{a}\right)$,
and $\hat{D}\left(\frac{s}{2}\right)=\exp\left[\frac{s}{2}\left(\hat{a}^{\dagger}-\hat{a}\right)\right]$.
The parameter $s$ is the ratio between the coupling $g$ and beam
width $\sigma$, i.e., $s=\frac{g}{\sigma}$. It takes continuous
value and could characterize the measurement strength. If $0<s<1$($s>1$),
the measurement is called weak (strong) measurement. For the implementation
of the postselected von Neumann mesaurement, the state $\vert\psi_{f}\rangle=\vert\uparrow_{z}\rangle$
is taken as the postselecting state onto the Eq.(\ref{eq:3}), then
the normalized final state of the pointer can be obtained as

\begin{equation}
\vert\Phi\rangle=\lambda\left[\!\left(1+\langle\hat{\sigma}_{x}\rangle_{w}\right)\hat{D}\left(\frac{s}{2}\right)+\!\!\left(1-\!\!\langle\hat{\sigma}_{x}\rangle_{w}\right)\hat{D}\left(\!-\frac{s}{2}\right)\right]\!\!\vert\phi\rangle,\label{eq:4}
\end{equation}
where the normalization coefficient $\lambda$ is defined as

\begin{align*}
\lambda & =\frac{1}{\sqrt{2}}[1+\!\vert\langle\hat{\sigma}_{x}\rangle_{w}\vert^{2}+(1\!-\!\vert\langle\hat{\sigma}_{x}\rangle_{w}\vert^{2})(1\!-\vert\beta\vert^{2})e^{-\frac{1}{2}\vert\beta\vert^{2}}]^{-\frac{1}{2}}
\end{align*}
with $\beta=-s[\cosh(r)-e^{i\theta}\sinh(r)].$ As the result of postselected
mesaurement, the weak value of system observable $\hat{\sigma_{x}}$
is given by 
\begin{equation}
\langle\hat{\sigma}_{x}\rangle_{w}=\frac{\langle\psi_{f}\vert\hat{\sigma}_{x}\vert\psi_{i}\rangle}{\langle\psi_{f}\vert\psi_{i}\rangle}=e^{i\delta}\tan\frac{\alpha}{2}.
\end{equation}
The Eq.(\ref{eq:4}) is the final state of the pointer after postselected
von Neumann measurement, which will be used throughout our work. It
can be easily seen that the weak value above can go beyond the scope
of the normal values of observable $\sigma_{x}$ and even takes a
complex value if the case of $\delta\neq0$ is taken into account.
As mentioned in the introduction part, the anomalous large weak values
can not only be used to amplify the tiny system information and also
be used to optimize the quantum states. Next we examine the effects
of anomalous values of measured system observable on the inherent
properties of SPSSVS.

\subsection{Squeezing parameter\label{sec:2}}

In this subsection, we study the effects of postselected von Nuemann
measurement on the squeezing parameter of SPSSVS. Squeezing effects
plays a crucial role in quantum theory and its applications. Squeezing
refers to the case where the quantum fluctuation in a state is less
than the vacuum fluctuation. Thus, the states with squeezing effects
are considered to be genuine nonclassical states \citep{PhysRevLett.103.213603},
and there is no classical counterpart \citep{Brooks2012NonclassicalLG,PhysRevLett.57.2520}
of them. Research on squeezing effects, especially on orthogonal squeezing
effects, has been conducted in optical communications and information
theory \citep{PhysRevLett.80.869,RevModPhys.58.1001,1056132,PhysRevA.72.053806,PhysRevA.76.011804,PhysRevD.23.1693},
gravitational wave detection \citep{PhysRevA.54.2614}, intensive
encoding \citep{PhysRevA.61.042302}, resonance fluorescence \citep{PhysRevA.72.053812}
and quantum cryptography \citep{PhysRevA.60.910} etc.

\begin{figure}
\begin{centering}
\subfloat[\label{fig:1.1a}]{\centering{}\includegraphics[width=8cm]{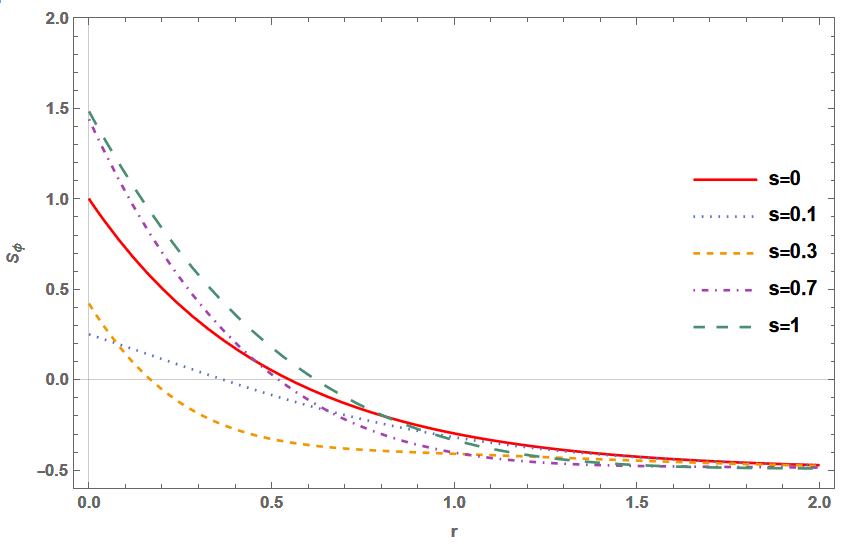}}
\par\end{centering}
\begin{centering}
\subfloat[\label{fig:1.1b}]{\centering{}\includegraphics[width=8cm]{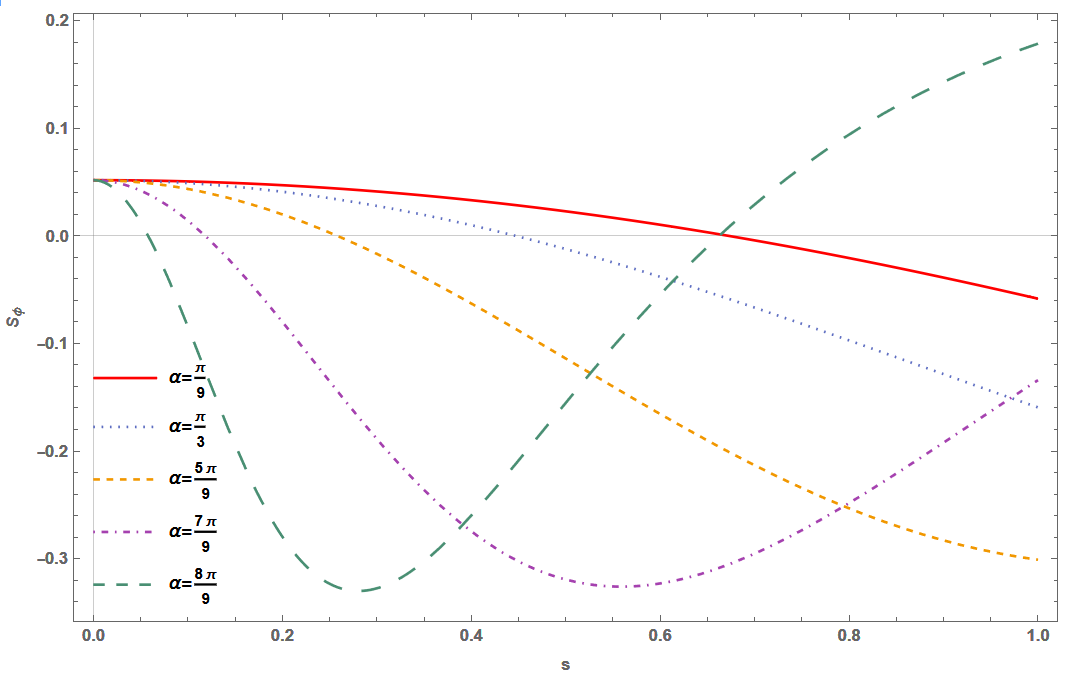}}
\par\end{centering}
\centering{}\subfloat[\label{fig:1.1c}]{\centering{}\includegraphics[width=8cm]{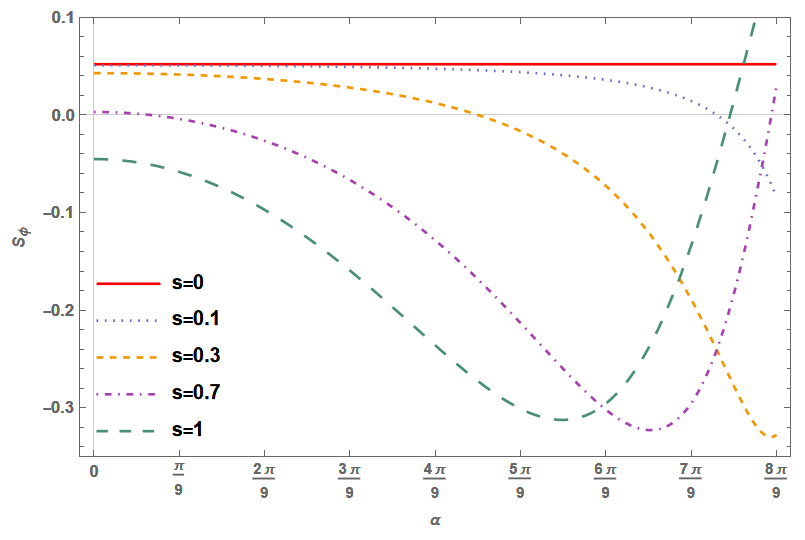}}\caption{\label{fig:1-1} Squeezing Parameter of SPSSVS under $\vert\Phi\rangle$
for different system parameters. (a) $S_{\phi}$ as a function of
$r$ for different coupling strength parameter $s$ while fixed the
weak value parameter $\alpha=\frac{8\pi}{9}$. (b) $S_{\phi}$ as
a function of $s$ for different weak values while fixed squeezing
state parameter $r=0.5$. (c) $S_{\phi}$ as a function of weak values
characterized by $\alpha$ for different $s$ while keeping $r=0.5$.
Here, we take $\theta=0,\delta=\phi=\frac{\pi}{2}$}
\end{figure}

The squeezing parameter of a radiation field is defined as
\begin{equation}
\hat{S}_{\phi}=(\triangle\hat{X}_{\phi})^{2}-\frac{1}{2},\label{eq:6}
\end{equation}
where 
\begin{equation}
\hat{X}_{\phi}=\frac{1}{\sqrt{2}}(\hat{a}e^{-i\phi}+\hat{a}^{\dagger}e^{i\phi}).\label{eq:7}
\end{equation}
is a quadrature operator of the field and satisfies $[\hat{X}_{\phi},\hat{X}_{\phi+\frac{\pi}{2}}]=i$,
$\phi=0,\frac{\pi}{2}$. The variance of $\hat{X}_{\phi}$ is defined
as $(\triangle\hat{X}_{\phi})^{2}=\langle\Phi\vert\hat{X}_{\phi}^{2}\vert\Phi\rangle-\langle\Phi\vert\hat{X}_{\phi}\vert\Phi\rangle^{2}$
under the state $\vert\Phi\rangle$. From the definition of $\hat{S}_{\phi}$,
it can be seen that $\hat{S}_{\phi}\ge-\frac{1}{2}$. If $-\frac{1}{2}\le\hat{S}_{\phi}\le0$,
then there occurs squeezing effect of corresponding radiation field
in phase space along $\phi=0$ or $\frac{\pi}{2}$ direction. Next
we calculate the above quantities under the state $\vert\Phi\rangle$
which is given in Eq.(\ref{eq:4}). By a straightforward calculation
we can get that

\begin{align}
\langle\hat{X}_{\phi}\rangle & =\frac{1}{\sqrt{2}}\langle\hat{a}e^{-i\phi}+\hat{a}^{\dagger}e^{i\phi}\rangle=\sqrt{2}\Re\left[\langle\hat{a}\rangle e^{-i\phi}\right]\label{eq:8}
\end{align}
 and

\begin{align}
\langle\hat{X}_{\phi}^{2}\rangle & =\frac{1}{2}\langle(\hat{a}e^{-i\phi}+\hat{a}^{\dagger}e^{i\phi})(\hat{a}e^{-i\phi}+\hat{a}^{\dagger}e^{i\phi})\rangle\nonumber \\
 & =\Re\left[\langle\hat{a}^{2}\rangle e^{-2i\phi}\right]+\langle\hat{a}^{\dagger}\hat{a}\rangle+\frac{1}{2}.\label{eq:9}
\end{align}
The explicit expressions of $\langle\hat{X}_{\phi}\rangle$ and $\langle\hat{X}_{\phi}^{2}\rangle$
can be obtained by calculating the expectaion values of $\langle\hat{a}\rangle$,$\langle\hat{a}^{2}\rangle$and
$\langle\hat{a}^{\dagger}\hat{a}\rangle$ under the state $\vert\Phi\rangle$
which are given as

\begin{align}
\langle\hat{a}\rangle & =\vert\lambda\vert^{2}[2\Re[\langle\hat{\sigma}_{x}\rangle_{w}]s-2i\Im[\langle\hat{\sigma}_{x}\rangle_{w}]h_{1}(s)],\label{eq:10}\\
\langle\hat{a}^{2}\rangle & =\vert\lambda\vert^{2}[(1+\vert\langle\hat{\sigma}_{x}\rangle_{w}\vert^{2})(3e^{i\theta}\sinh(2r)+\frac{s^{2}}{2}),\nonumber \\
 & +2(1-\vert\langle\hat{\sigma}_{x}\rangle_{w}\vert^{2})h_{2}(s)],\label{eq:11}
\end{align}
 and 
\begin{align}
\langle\hat{a}^{\dagger}\hat{a}\rangle & =2\vert\lambda\vert^{2}[(1+\vert\langle\hat{\sigma}_{x}\rangle_{w}\vert^{2})(1+3\sinh^{2}(r)+\frac{s^{2}}{4})\nonumber \\
 & +(1-\vert\langle\hat{\sigma}_{x}\rangle_{w}\vert^{2})h_{3}(s)],\label{eq:12}
\end{align}
respectively. Here,

\begin{align}
h_{1}(s) & =[\beta\cosh(r)(\beta^{2}e^{-i\theta}\coth(r)+3)\nonumber \\
 & +\frac{s}{2}(2\beta^{2}e^{-i\theta}\coth(r)-\vert\beta\vert^{2}+3)]e^{-\frac{1}{2}\vert\beta\vert^{2}},\label{eq:13}
\end{align}

\begin{align}
h_{2}(s) & =\{\beta^{2}\cosh^{2}(r)[\beta^{2}e^{-i\theta}\coth(r)+6]+\frac{3}{2}e^{i\theta}\sinh(2r)\nonumber \\
 & -2s\beta\cosh(r)(\beta^{2}e^{-i\theta}\coth(r)+3)\nonumber \\
 & +\frac{s^{2}}{4}[2\beta^{2}e^{-i\theta}\coth(r)-\vert\beta\vert^{2}+3]\}e^{-\frac{1}{2}\vert\beta\vert^{2}},\label{eq:14}
\end{align}
 and

\begin{align}
h_{3}(s) & =\{\beta^{2}e^{-i\theta}[\coth(r)\cosh^{2}(r)+5\sinh(r)\cosh(r)\nonumber \\
 & +\beta^{2}e^{-i\theta}\cosh^{2}(r)]+1+3\sinh^{2}(r)\nonumber \\
 & +\frac{3s}{2}\beta e^{-i\theta}[\frac{1+3\sinh^{2}(r)}{\sinh(r)}+\beta^{2}e^{-i\theta}\cosh(r)]\nonumber \\
 & +\!\!\frac{s^{2}}{2}e^{-i\theta}[\coth(r)+\!\beta^{2}e^{-i\theta}]\!+\frac{se^{-i\theta}}{2}\beta^{3}\cosh(r)\coth(r)\nonumber \\
 & \!\!+\!\frac{3s}{2}\beta\cosh(r)\!+\frac{s^{2}}{4}[2\beta^{2}e^{-i\theta}\coth(r)\!-\!\vert\beta\vert^{2}\!+\!3]\}e^{-\frac{1}{2}\vert\beta\vert^{2}}.\label{eq:15}
\end{align}
 By substituting the above expressions into Eq. (\ref{eq:6}) we can
obtain the squeezing parameter of SPSSVS under the state $\vert\Phi\rangle$.
Since its explicit expression is too cumbersome to analytically analyze,
we give numerical results. The numerical results for different parameters
are shown in Fig.\ref{fig:1-1}. In Fig.\ref{fig:1.1a}, we presented
the $\hat{S}_{\pi/2}$ as a function of squeezing parameter $r$ for
different coupling strength $s$ while the weak value is fixed to
$\langle\hat{\sigma}_{x}\rangle_{w}=5.76$ ($\alpha=8\pi/9$). As
we can see in the weak measurement regime $(0<s\ll1)$, the squeezing
parameter $\hat{S}_{\pi/2}$ takes lower values than the initial state
$\vert\phi\rangle$. In Fig.\ref{fig:1.1b} the dependence of $\hat{S}_{\pi/2}$
on the coupling strength $s$ for different anomalous weak values
and fixed squeezing parameter $r$=0.5. The larger the anomalous weak
values, the larger the squeezing parameters. The squeezing parameter
$\hat{S}_{\pi/2}$ as a function of anomalous weak values is plotted
in Fig.\ref{fig:1.1c} for different $s$. All these results further
confirm that the squeezing effect of SPSSVS is increased after taking
weak measurement.

\subsection{Mandel factor\label{sec:2-1}}

The photon number statistics of a radiant light field typically exhibit
three types of distributions: super-Poissonian, Poisson, and sub-Poissonian.
The Poisson distribution is the statistical distribution of a beam
of perfectly coherent light with constant intensity, and serves as
a benchmark for classifying statistical distributions. A distribution
that is broader than the Poisson distribution is referred to as a
super-Poissonian distribution, while a distribution that is narrower
than the Poisson distribution is referred to as a sub-Poissonian distribution.
The sub-Poissonian distribution exhibits lower photon number fluctuations
compared to the Poisson distribution, and lacks a classical counterpart.
Therefore, it is important to describe the state using non-classical
terminology with and mandel factor. In this subsection, we investigate
the Mandel factor $Q_{m}$ for the $\vert\Phi\rangle$ state.

The mathematical definition of the Mandel factor $Q_{m}$of a single-mode
radiation field is given as

\begin{equation}
Q_{m}=\frac{\langle\hat{a}^{\dagger2}\hat{a}^{2}\rangle}{\langle\hat{a}^{\dagger}\hat{a}\rangle}-\langle\hat{a}^{\dagger}\hat{a}\rangle.\label{eq:16}
\end{equation}
If $-1\le Q_{m}<0$, the field exhibits sub-Poissonian statistics
and is purely non-classical, and the Mandel factor can never be less
than $-1$ for any cases \citep{Agarwal2013}.

\begin{figure}
\begin{centering}
\includegraphics[width=8cm]{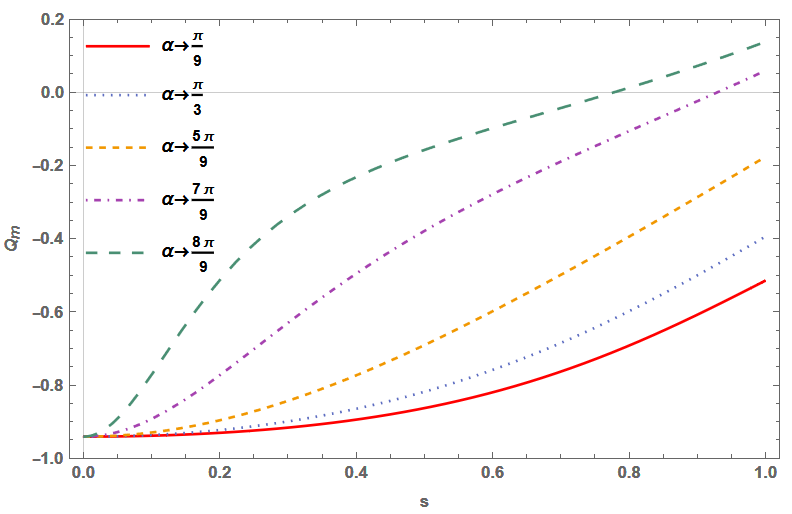}
\par\end{centering}
\caption{Mandel Factor $Q_{m}$of the SPSSVS after psotselected measurement.
Here $\ensuremath{\theta}=0,\ensuremath{\delta}=\frac{\pi}{2},r=0.1$\label{fig:2 mandel fuctor}}
\end{figure}

In Fig.\ref{fig:2 mandel fuctor} the variation of $Q_{m}$ with the
coupling strength $s$ is plotted for the fixed value of $\alpha$.
It is evident that, as the weak value increases, the nonclassical
nature weakens.

It is also observed that the nonclassical properties weaken as s shifts
from the weak measurement to the strong one. We noted that for postselected
von Neumann measurements based on SPSSVS, the value of $Q_{m}$ decreases
significantly with very large weak values. This is contrary to the
signal amplification feature verified in previous studies\citep{PhysRevA.92.022109,PhysRevLett.115.120401,Turek_2020}.

\subsection{Wigner function \label{sec:4}}

In order to deeply understand the effects of postselected von Neumann
measurement on the inherent properties of SPSSVS, in this subsection,
we examine the phase space distribution of SPSSVS by calculating the
Wigner function. Wigner function contains the quasi-probability distribution
of the quantum state, phase, and other important information, similar
to the wave function or density matrix corresponding to a quantum
state. A negative value of Wigner function of a given state means
the nonclassicality of that state. The bigger negative region in phase
space a state has, the more nonclassicality it has. In order to obtain
even more information about the state $\vert\Phi\rangle$, it is necessary
to investigate the Wigner function in phase space . In general, the
Wigner function is defined as the two-dimensional Fourier transform
of the symmetric order eigenfunction. The general expression of Wigner
function for a state $\rho=\vert\Phi\rangle\langle\Phi\vert$ is given
by \citep{Int}

\begin{equation}
W(z)\equiv\dfrac{1}{\pi^{2}}\int_{-\infty}^{+\infty}\exp(\lambda^{*}z-\lambda z^{*})C_{W}(\lambda)d^{2}\lambda.
\end{equation}
where $C_{W}(\lambda)$ is the symmetrically ordered characteristic
function defined as

\begin{equation}
C_{W}(\lambda)=Tr\left[\rho e^{\lambda\hat{a}^{\dagger}-\lambda^{*}\hat{a}}\right].
\end{equation}
Here, we use the notation of $\lambda^{\prime}$ and $\lambda^{\prime\prime}$
for the real and imaginary parts of $\lambda$ and set $z=x+ip$ to
emphasize the analogy between the quadratic radiation field and the
normalized dimensionless position and momentum observables of the
beam in phase space. We can rewrite the definition of the Wigner function
in terms of $x,p$ and $\lambda^{\prime},\lambda^{\prime\prime}$
as

\begin{equation}
W(x,p)=\frac{1}{\pi^{2}}\int e^{2i(p\lambda^{\prime}-x\lambda^{\prime\prime})}C_{W}(\lambda)d\lambda^{\prime}d\lambda^{\prime\prime}.\label{eq:wigner-function}
\end{equation}

By substiting the final normalized pointer state $\vert\Phi\rangle$
into Eq.(\ref{eq:wigner-function}), the explicit expression of its
Wigner function can be calculated as following

\begin{align}
W(z) & =\frac{1}{\pi^{2}}\int\int d\lambda d\lambda^{\ast}Tr\left[\rho D(\lambda)\right]e^{-(\lambda\alpha^{\ast}-\lambda^{\ast}\alpha)}\nonumber \\
 & =\vert\lambda\vert^{2}[\vert1+\langle\hat{\sigma}_{x}\rangle_{w}\vert^{2}\frac{2}{\pi}(4|\alpha^{\prime}-s\Gamma|^{2}-1)\mathrm{e}^{-2|\alpha^{\prime}-s\Gamma|^{2}}\nonumber \\
 & +\vert1-\langle\hat{\sigma}_{x}\rangle_{w}\vert^{2}\frac{2}{\pi}(4|\alpha^{\prime}+s\Gamma|^{2}-1)\mathrm{e}^{-2|\alpha^{\prime}+s\Gamma|^{2}}\nonumber \\
 & +\frac{4}{\pi}\Re\left[(1+\langle\hat{\sigma}_{x}\rangle_{w})(1-\langle\hat{\sigma}_{x}\rangle_{w})^{\ast}e^{-2s(\alpha-\alpha^{*})-2|\alpha^{\prime}|^{2}}\right]\nonumber \\
 & \times(4|\alpha^{\prime}|^{2}-1)],\label{eq:20}
\end{align}
where $\Gamma=\cosh(r)-e^{i\theta}\sinh(r)$and $\alpha^{\prime}=\alpha\cosh(r)-\alpha^{*}e^{i\theta}\sinh(r)$.
In general, this is a real Wigner function and its value is bounded
as $-\frac{2}{\pi}\leq W(z)\leq\frac{2}{\pi}$ in whole phase space
.

\begin{widetext}

\begin{figure}
\begin{centering}
\includegraphics[width=15cm,height=15cm]{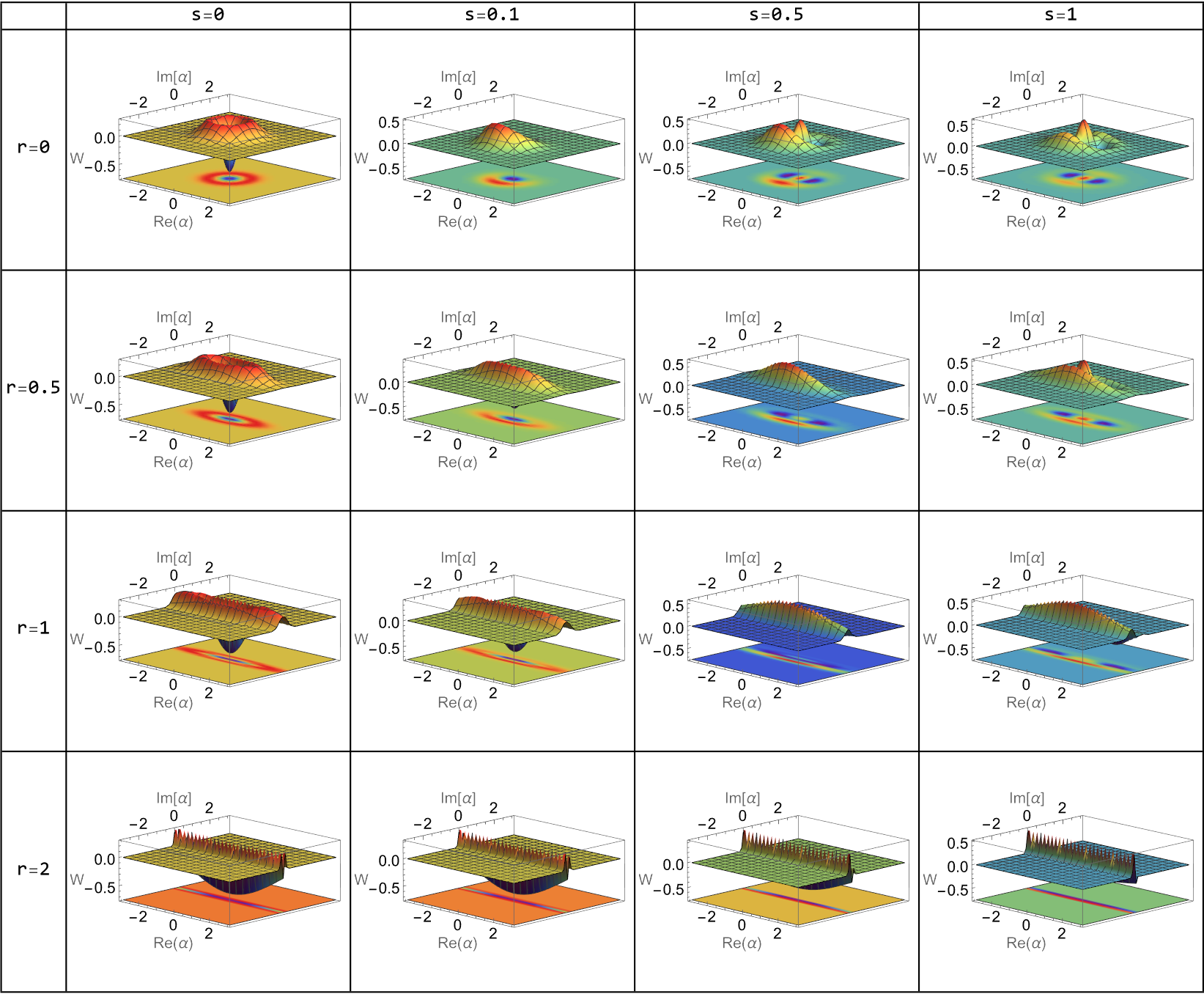}
\par\end{centering}
\caption{(Color Online)\label{fig:Wigner-f} The Wigner function of SPSSVS
after postselected von Neumann measurement. Here, we take $\ensuremath{\theta}=0,\ensuremath{\delta}=\frac{\pi}{2},\ensuremath{\alpha=\frac{8\pi}{9}}$.
Each column represents different measurement strengths with s=0,0.1,0.5,1,
and are ordered accordingly from left to right.}
\end{figure}

\end{widetext}

To show the effects of the postselected von Neumann measurement on
the non-classical feature of SPSSVS, in Fig.\ref{fig:Wigner-f} we
plot the Wigner function of SPSSVS for different squeezing parameters
$r$ and coupling strengths $s$. Each column from left to right indicates
the different coupling strengths $s$ for 0, 0.1,0.5 and 1, and each
row from up to down represents the different squeezing parameters
$r$ for $0,0.5,1$ and $2$. Here, we take the weak value as $\langle\hat{\sigma}_{x}\rangle_{w}=5.76$
($\alpha=8\pi/9$). It is observed that as showed in squeezing parameter
$\hat{S}_{\pi/2}$, there always occur squeezing effects along $y$-axis,
which are increased in weak measurement regimes. As is shown, the
positive peaks of the Wigner function gradually becomes irregular
with increasing the coupling strength $s$. Furthermore, in Fig.\ref{fig:Wigner-f}
for $s=0.5,1$ and $r=0.5,1$ cases we can see that significant interference
structures manifest and the negative regions become larger than the
initial pointer state.

As mentioned above, the existence of and progressively stronger negative
regions of the Wigner function in phase space indicates the degree
of nonclassicality of the associated state. From the above analysis
we can conclude that after the postselected von Neumann measurement,
the phase space distribution of SPSSVS is not only more squeezed but
its nonclassicality also gets more obvious in the most regimes.

\section{Signal-to-noise ratio\label{sec:5}}

In precision measurement it is important to get precise information
whild suppress the associated noise. To demonstrate the superiority
of SPSSVS-based postselected measurement over non-postselected measurement\citep{Agarwal2013}
for position shift, the SNR between postselected and non-postselected
measurement is analyzed, which is characterized as below 
\begin{equation}
\chi=\frac{\mathcal{R}_{p}}{\mathcal{R}_{n}}.
\end{equation}
where$\mathcal{R}_{p}$ represents the SNR of postselected von Neumann
measurement which is defined by

\begin{equation}
\mathcal{R}_{p}=\frac{\sqrt{NP_{s}}\vert\delta x\vert}{\triangle x}
\end{equation}
with the variance of position operator

\begin{equation}
\triangle x=\sqrt{\langle\Phi|\hat{X}^{2}|\Phi\rangle-\langle\Phi|\hat{X}|\Phi\rangle^{2}},
\end{equation}
and the average shift of the pointer variable $x$ after postselected
measurement

\begin{equation}
\delta x=\langle\Phi|\hat{X}|\Phi\rangle-\langle\phi|\hat{X}|\phi\rangle.
\end{equation}
Here, $\hat{X}=\sigma\left(\hat{a}+\hat{a}^{\dagger}\right)$ is the
position operator, $N$ is the total number of measurements, and $P_{s}=\vert\langle\psi_{f}\vert\psi_{i}\rangle\vert^{2}=\cos^{2}\frac{\alpha}{2}$
is the success probability of postselection, and $\vert\Phi\rangle$
denotes the normalized state of SPSSVS after postselection which is
given by Eq.(\ref{eq:4}). We know that

\begin{align}
\langle\phi\vert\hat{X}\vert\phi\rangle & =2\sigma\Re\left[\langle\phi\vert\hat{a}\vert\phi\rangle\right]\\
\langle\Phi\vert\hat{X}\vert\Phi\rangle & =2\sigma\Re\left[\langle\Phi\vert\hat{a}\vert\Phi\rangle\right]\\
\langle\Phi\vert\hat{X}^{2}\vert\Phi\rangle & =\sigma^{2}(2\langle\Phi\vert\hat{a}^{\dagger}\hat{a}\vert\Phi\rangle+2Re[\langle\Phi\vert\hat{a}^{2}\vert\Phi\rangle]+1)\label{eq:27}
\end{align}
Furthermore, the SNR $\mathcal{R}_{n}$ for non-postselected measurement
is defined as

\begin{equation}
\mathcal{R}_{n}=\frac{\sqrt{N}\vert\delta x^{\prime}\vert}{\triangle x^{\prime}}
\end{equation}
 with

\begin{align}
\triangle x' & =\sqrt{\langle\Psi|\hat{X}^{2}|\Psi\rangle-\langle\Psi|\hat{X}|\Psi\rangle^{2}},\nonumber \\
\delta x' & =\langle\Psi|\hat{X}|\Psi\rangle-\langle\phi|\hat{X}|\phi\rangle.
\end{align}
Here, we have to note that $\vert\Psi\rangle$ is the normalized final
state of the total systemwithout taking postselection as is given
in Eq.(\ref{eq:3}).

The expectation values of $\langle\hat{a}\rangle$, $\langle\hat{a}^{\dagger}\hat{a}\rangle$
and $\langle\hat{a}^{2}\rangle$ under the state $\vert\Psi\rangle$
are given as

\begin{align}
\langle\Psi\vert\hat{a}\vert\Psi\rangle & =\frac{s}{2}\sin\alpha\cos\delta,\\
\langle\Psi\vert\hat{a}^{\dagger}\hat{a}\vert\Psi\rangle & =1+3\sinh^{2}(r)+\frac{s^{2}}{4},
\end{align}
 and

\begin{equation}
\langle\Psi\vert\hat{a}^{2}\vert\Psi\rangle=\frac{1}{2}(3e^{i\theta}\sinh(2r)+\frac{s^{2}}{4}),\label{eq:32}
\end{equation}
 respectively.

\begin{figure}
\begin{centering}
\subfloat[\label{fig:SNR-A}]{\begin{centering}
\includegraphics[width=8cm]{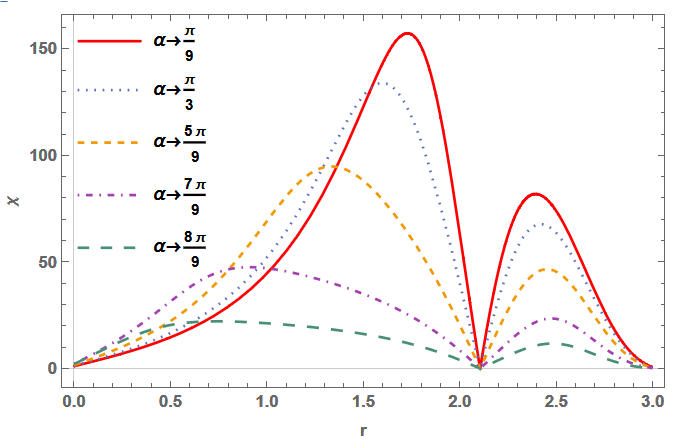}
\par\end{centering}
}
\par\end{centering}
\begin{centering}
\subfloat[\label{fig:SNR-B}]{\begin{centering}
\includegraphics[width=8cm]{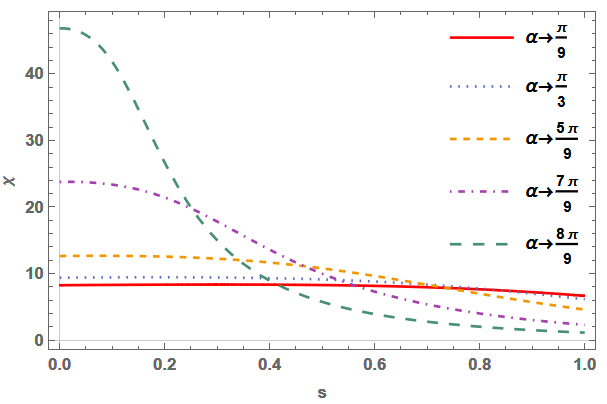}
\par\end{centering}
}
\par\end{centering}
\caption{(Color online ) The ratio $\chi$ of SNRs of SPSSVS between postselected
and non-postselected measurement cases . Here, we take $\theta=\frac{5\pi}{12},\delta=\frac{\pi}{2}$,
In (a), $\chi$ is plotted as a function of squeezing parameter $r$
for different weak values while fixed $s=0.3$. In (b), $\chi$ is
plotted as a function of coupling strength parameter $s$ for different
weak values for the fixed coupling paramter $r=0.3$.\label{fig:SNR }}
\end{figure}

In Fig.\ref{fig:SNR }, we show the ratio $\chi$ of SNRs between
postselected and nonpostselected von Neumann measurement for different
system parameters. As presented in Fig.\ref{fig:SNR-A}, we plot the
ratio $\chi$ as a function of the coherent state parameter $r$ and
we can see that in the weak measurement regime the ratio $\chi$ of
SNRs exhibits a slightly damping periodic oscillation with the incrasing
of system parameter $r$, and in the most of the regions, the ratio
$\chi$ is much larger than one. Furthermore, as indicated in Fig.\ref{fig:SNR-B},
the ratio $\chi$ increases in weak measurement regime ($s\ll1$)
for large anamolous weak values. In a word, it can be found that the
signal-to-noise ratio of SPSSVS aftewhich isr postselected measurement
is increased dramaticaly in weak measurement regime for large weak
values compared to nonpostselected measurement case. This result indicates
that SPSSVS may have potential applications in the associated presicion
measurement problems after taking postselected weak meaurement procedure.

\section{STATE DISTANCE\label{sec:6}}

As discussed in above context, the inherent properties of SPSSVS are
changed dramatically after taking postselected von Neumann measurement.
As shown in previous studies, the postselected measurement could change
the state, and it may also be a reason to change the nonclassical
features of original given state. In this section, to examine the
effects of postselected measurement on the initial state $\vert\phi\rangle$,
we check the state distance. In quantum information theory the distance
between two quantum states described by density operators $\rho$
and $\sigma$, is defined as

\begin{equation}
F=\left(Tr\sqrt{\sqrt{\rho}\sigma\sqrt{\rho}}\right)^{2}.
\end{equation}

This formula also can characterize the fidelity (and is also called
Uhlmann-Jozsa fidelity) between two states. In our present study both
states are pure i.e., $\rho=\vert\phi\rangle\langle\phi\vert$ and
$\sigma=\vert\Phi\rangle\langle\Phi\vert$, then the above formula
can be rewritten as

\begin{equation}
F=|\langle\phi|\Phi\rangle|^{2}.
\end{equation}
Its value is bounded $0\leq F\leq1$. If $F=1(F=0)$, then the two
states are exactly the same (totally different). This quantity is
a natural candidate for the state distance because it corresponds
to the closeness of states in the Hilbert space. The inner product
of $\vert\phi\rangle$ and $\vert\Phi\rangle$ is calculated as

\begin{align}
\langle\phi|\Phi\rangle & =\lambda\left[(1+\langle\hat{\sigma}_{x}\rangle)P_{1}(s)+(1-\langle\hat{\sigma}_{x}\rangle)P_{2}(s)\right],
\end{align}
 where

\begin{align}
P_{1}(s) & =[\beta^{2}e^{-i\theta}\coth(r)+1]e^{-\frac{1}{2}\vert\beta\vert^{2}}-\frac{s\beta e^{-i\theta}}{2\sinh(r)},
\end{align}
 and 
\begin{equation}
P_{2}(s)=[\beta^{2}e^{-i\theta}\coth(r)+1]e^{-\frac{1}{2}\vert\beta\vert^{2}}+\frac{s\beta e^{-i\theta}}{2\sinh(r)}.\label{eq:37}
\end{equation}

\begin{figure}
\begin{centering}
\includegraphics[width=8cm]{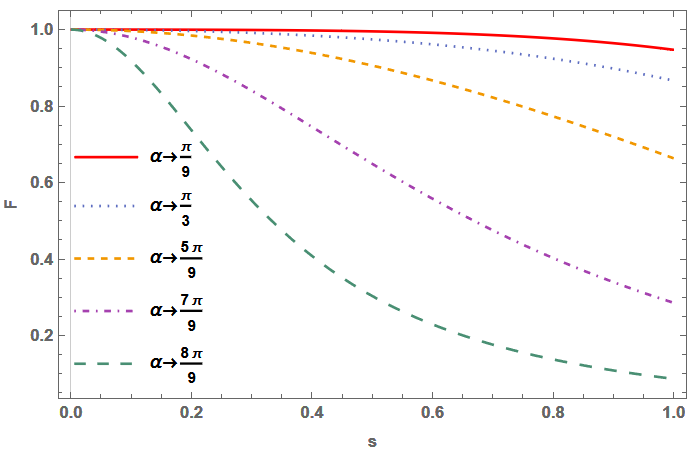}
\par\end{centering}
\caption{(color online) The state distance between $\vert\phi\rangle$ and
$\vert\Phi\rangle$.Here we take $r=0.5,\theta=0,\delta=\frac{\pi}{2}$.\label{fig:5}}
\end{figure}

In Fig.\ref{fig:5}, the fidelity $F$ is plotted as a function of
coupling strength $s$ for the fixed $r=0.5$. As showed in Fig.\ref{fig:5},
the fidelity between $\vert\phi\rangle$ and $\vert\Phi\rangle$ is
gradually decreased with increasing the coupling strength and weak
values. This indicates that the postselected von Neumann measurement
can change the inherent properties of initial given state by destroying
the given state gradually and change it to another state as some quantum
state engineering processes.

\section{CONCLUSION AND REMARKS\label{sec:7}}

In summary, we investigated the effects of postselected von Neumann
measurement on the properties of SPSSVS. We firstly presented both
the analytical and numercial results of Mandel factors, squeezing
parameter, and Wigner function of final pointer states. It is found
that in weak measurement regime the squeezing effect is increased
dramatically for anomalous weak values. We also noticed that the postselected
measurement has no positive effect to the Mandel factor of SPSSVS.
By investigating the Wigner function we found that in weak measurment
regime the negativity of Wigner function of SPSSVS is increased compared
to the initial state. These results showed that postselected von Neumann
measurement can increase the nonclassicality of SPSSVS in weak measurement
regime by properly choosing the anomalous weak values. Furthermore,
the comparison of SNRs between postselected and non-postselected measurement
cases showed that the SNR of postselected measurement scheme is much
larger than non-postselected case. We noticed that in this process
the amplification effect of weak value played a significant role although
the postselection probablity is low in this case.

To conclude, in our current work, we investigated the effects of postselected
von-Neuman measurement on the inherent properties of the SPSSVS. However,
real light beams are time-dependent, and representing this dependence
needs two or more modes of the optical systems. Therefore, it would
be interesting to investigate the impact of postselected von-Neumann
measurements on other relevant radiation fields, such as pair-coherent
state (two-mode radiation field)\citep{PhysRevA.50.2865,PhysRevA.13.2226}
and other multimode radiation fields \citep{PhysRevA.42.4102}as well.
\begin{acknowledgments}
This work was supported by the National Natural Science Foundation
of China (No. 12365005) and (No. 11864042).
\end{acknowledgments}

\bibliographystyle{apsrev4-1}
\bibliography{ref11}

\end{document}